# Weak vs. Strong Correlations: Bid-Ask Spreads for Weather-Contingent Options

René Carmona[#, 1, 2, 3] and Dario Villani[*, 4]

[1] Department of Operations Research and Financial Engineering
Princeton University, Princeton, NJ 08544

[2] Bendheim Center for Finance
Princeton University, Princeton, NJ 08544

[3] Applied and Computational Mathematics Program
Princeton University, Princeton, NJ 08544

[4] Hess Energy Trading Company, LLC
1185 Avenue of the Americas, New York, NY 10036

ABSTRACT

We price weather-contingent options by use of Monte Carlo simulations. After calibrating the models to fit quoted prices, we analyze bid-ask spreads in terms of correlations across markets. Results are presented for a double-trigger Weather vs. Natural Gas call option.

[#] R. Carmona is the Paul. M. Wythes '55 Professor of Engineering and Finance (e-mail: rcarmona@princeton.edu)
[*] D. Villani is a Commodity Trader (e-mail: dvillani@alumni.princeton.edu)

Weather-contingent options are present in most commodity markets [1] for the following reasons. First, they offer an inexpensive way of having long volatility positions in commodity markets where the skew is positive (i.e., investors are scared of prices gapping up). Second, they transfer liquidity from a more efficient market to a less efficient one. This is the case for the natural gas and weather markets which we consider in this paper.

Despite these two significant reasons, the market for weather-contingent options is still far from being liquid. This occurrence can be ascribed to the lack of sound evaluation methods that could help hedging on the side of market makers and speculators, and price discovery on the side of end-users.

Our working hypothesis is that the temperature $T$ and the commodity price $G$ are stochastic variables driven by two Wiener processes $W^{(T)}$ and $W^{(G)}$, respectively. That is,

$$T = T\left(\vec{\tau}, \sigma^{(T)} W^{(T)}\right)$$

and

$$G = G\left(\vec{\gamma}, \sigma^{(G)} W^{(G)}\right).$$

$\vec{\tau}$ and $\vec{\gamma}$ are parameters specific to a model dynamics. $\sigma^{(T)}$ is the volatility of the temperature innovations. $\sigma^{(G)}$ is the volatility of the commodity price innovations. The Wiener processes are correlated in the sense that they satisfy

$$dW_t^{(T)} dW_t^{(G)} = \rho dt.$$

The pricing scheme requires the temperature to be modeled with a stochastic process (see, e.g., Ref. [2]). Burn-cost methods cannot be formulated in terms of Wiener processes [3]; in order to introduce correlations in a meaningful way, we need to reach the same level of microscopic description usually used for commodity prices.

It is worth pointing out that the commodity price dependence on the temperature is restrictive in the present analysis. Indeed, the source of statistical correlation between the two stochastic variables $T$ and $G$ is limited to the noise terms driving the individual dynamics. The components of the vector $\vec{\gamma}$ could be functions of the temperature and so could be the volatility $\sigma^{(G)}$. A theoretical analysis of these models is possible, but the level of mathematical sophistication needed is such that the results loose their intuitive appeal. The interested reader is referred to Ref. [3] for details and further references.

Every day, the markets fix the value of the parameters $\vec{\tau}, \sigma^{(T)}, \vec{\gamma},$ and $\sigma^{(G)}$ for weather and the commodity separately. These values are usually obtained by proprietary blends of statistical estimation procedures applied to historical data, and calibration techniques applied to the prices of the active instruments. In this paper we assume that we have chosen our favorite method to risk adjust separately the two dynamics by calibration to the traded instruments. Then, we show how to combine these two calibrated univariate models into a bivariate model appropriate for the pricing of the double-trigger weather vs. natural gas call option. This will obviously involve the correlation coefficient $\rho$.

We consider the case (some of the data values are fabricated for the sake of simplicity) of a double-trigger weather vs. natural gas call option being priced on April

11, 2003. At the end of the contract period (August 1, 2003 through August 31, 2003), the seller pays the buyer for each day the average temperature $T_{avg}$ in New York is above $T_K = 84$ degrees Fahrenheit and the Daily Gas Daily Index (DGDI) exceeds the Monthly Gas Daily Index (MGDI). The payout $\Pi$ is obtained as the sum of the daily values $\max\{DGDI - MGDI, 0\}$ multiplied by the volume $V$. The volume is typically $10,000$ mmBtu. On any given day, $T_{avg}$ is the semi-sum of the high and the low for the day. The weather station is LaGuardia International Airport. DGDI and MGDI are for Henry Hub of Louisiana-Onshore South. We have

$$\Pi = V \sum_i \theta\left[T_{avg}(i) - T_K\right] \max\{DGDI(i) - MGDI, 0\}$$

where $i$ runs over the days in August. $\theta$ is equal to 1 for positive arguments and 0 otherwise. On April 11, 2003, the DGDI is $5 and the risk-free interest rate is $1\%$.

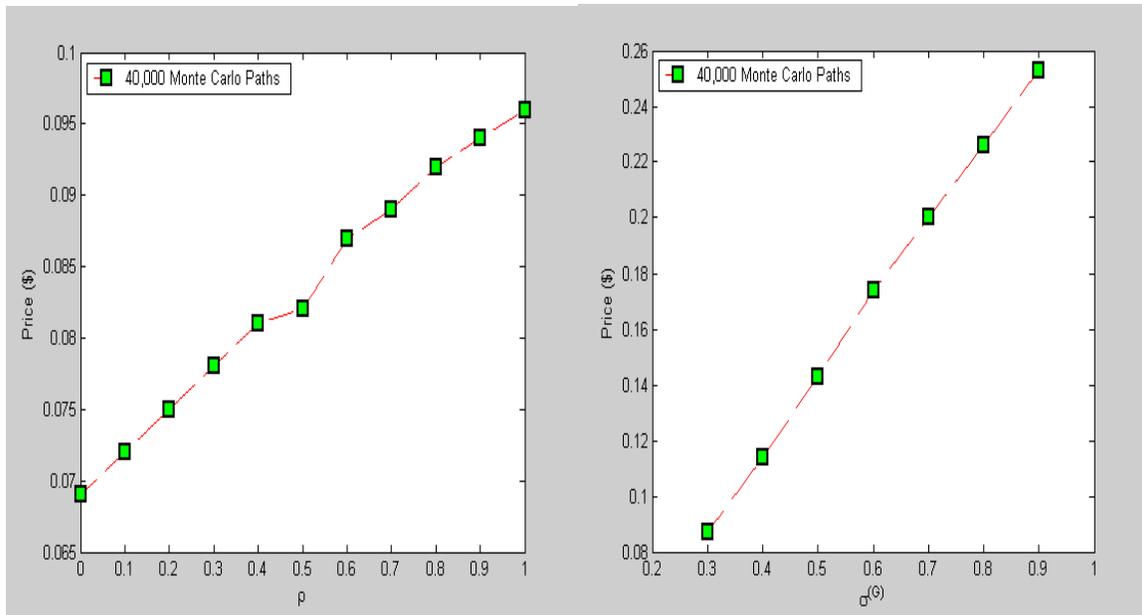

Figure 1. The price of the double-trigger weather vs. natural gas call option is plotted against the correlation coefficient $\rho$ (left panel) and the volatility $\sigma^{(G)}$ (right panel).

First, we calibrate the models by using the quotes available in the natural gas and weather markets. The market price of the natural gas option without the weather trigger is $0.36 per day and unit volume. The options market for weather derivatives implies a probability of the weather event $T_{avg} \geq 84$ approximately equal to $20\%$. For our choice of the underlying model dynamics, these values give the estimates $\sigma^{(G)} = 30\%$ and $\sigma^{(T)}$ equal to 1.5 times the 10-year historical volatility. Different models would imply different values of the volatilities. For example, the same market price for the natural gas option without the weather trigger can be obtained by use of a jump-diffusion model with a lower volatility and few jumps per year [3]. After the calibration is complete, we can run the simulation of a two dimensional stochastic process with only one degree of freedom,

$\rho$. Before proceeding further, it is worth mentioning that each numerical simulation in this paper has been done with 40,000 antithetic Monte Carlo paths.

In the left panel of Figure 1 we show how the price $\wp$ of the double-trigger weather vs. natural gas call option depends upon the correlation coefficient $\rho$. As expected, we find that for $\rho = 0$ (i.e., when weather and gas are driven by two independent Wiener processes) the price reduces to the one of the natural gas call option $0.36 multiplied by the market probability 20% of weather triggers ($\sim \$0.07$). This is a lower bound for the double-trigger option. One obvious, but not very constraining, upper bound is given by the price of the natural gas option $0.36. For positive correlations (an identical analysis can be carried out for negative correlations during the winter season), $\wp$ is an increasing function of $\rho$ with an upper bound at $\rho = 1$ that is 40% larger than the value at $\rho = 0$.

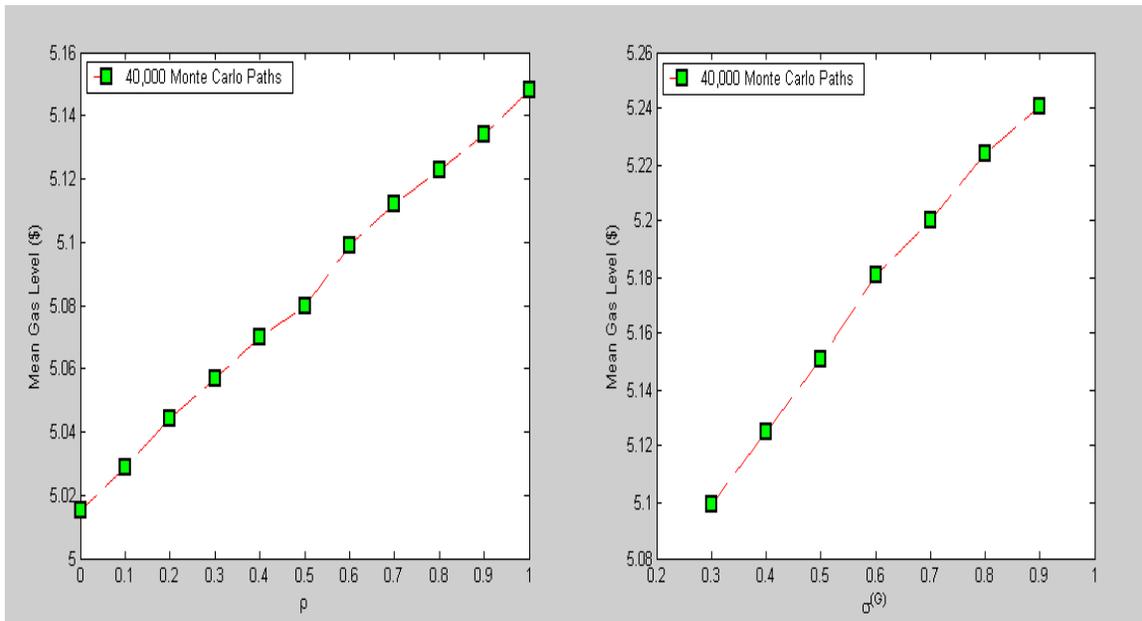

Figure 2. The mean level of gas price for $T_{avg} \geq 84$ is plotted against the correlation coefficient $\rho$ (left panel) and the volatility $\sigma^{(G)}$ (right panel).

So if all the other parameters (including the volatilities) are held fixed, the price is in one-to-one correspondence with the correlation coefficient $\rho$, and prices can be quoted in correlation coefficient units, in the same way European calls and puts are priced in terms of their implied volatilities.

At this point, a market player believes that a fair price for the double-trigger option needs to be in the range $0.069 – $0.096 given in the left panel of Figure 1. Very likely the same market player would go short the double-trigger option if another would offer at $0.1. Despite its obvious intuitive appeal, this type of analysis can be disastrous.

Even if the gas options market implies the volatility $\sigma^{(G)}$ equal to 30%, this is not a good estimate for the volatility during the days in which the weather triggers (i.e., when it is very hot in NY LaGuardia International Airport). Without going into the technical aspects of conditional variances, it is only fair to use increased gas volatilities to price double-trigger options. This is not a detail as we show in the right panel of Figure 2 where we fixed $\rho = 0.6$. In fact, as a function of the natural gas volatility, the price $\wp$ seems to grow at the rate of $0.3$. In this respect, if $\rho = 0$ represents the lower bound (starting bid level) for the option, the upper bound (starting ask level) could easily be $3-4$ times as much. In both panels of Figure 2 we show the same results in terms of mean level of the natural gas on the days for which the weather triggers. This is a complementary view where the price is not mapped on a value of the correlation but instead on the mean price of gas during hot days.

In conclusion, we have shown how to price weather-contingent options by use of Monte Carlo simulations. We have analyzed in some detail the double-trigger weather vs. natural gas call option. The correlation between the natural gas and weather markets emerged as a quoting device similar to the implied volatility of the Black-Scholes paradigm. Finally, it is worth pointing out that our approach gives exact analytical formula in some limiting regimes: more work in this direction is in progress.

## REFERENCES

[1] N. Ernst, *Bringing it all together*, Environmental Finance **4**, 28 (February 2003).
[2] F. Dornier, M. Queruel, *Caution to the wind*, EPRM, 30 (August 2000).
[3] R. Carmona, D. Villani, *Weather Derivatives*, Princeton Univ. Press (2004).